# Note on "XMM-Newton observations of the first unidentified TeV gamma-ray source TeV J2032+4130" by Horns et al. (astro-ph/0705.0009)

Y. Butt

*Harvard-Smithsonian Center for Astrophysics, 60 Garden St., Cambridge, MA 02138*

Horns et al (2007) report on XMM observations of TeV J2032+4130, where they find "**an extended X-ray emission region with a full width half maximum (FWHM) size of ≈ 12 arc min**" and flux ~7 ×$10^{-13}$ ergs $cm^{-2}$ $sec^{-1}$ (2-10 keV).

To their credit, they mention that the "**observed extended X-ray emission could in principle be a so far unknown population of faint X-ray sources that are by chance distributed at an angular size which is similar to the one of TeV J2032+4130**". Indeed, this is the more likely scenario – except that it is by no means "so far unknown". Such sources have been seen in the higher-resolution Chandra observations already reported by Butt et al. (2006) – Fig 1.

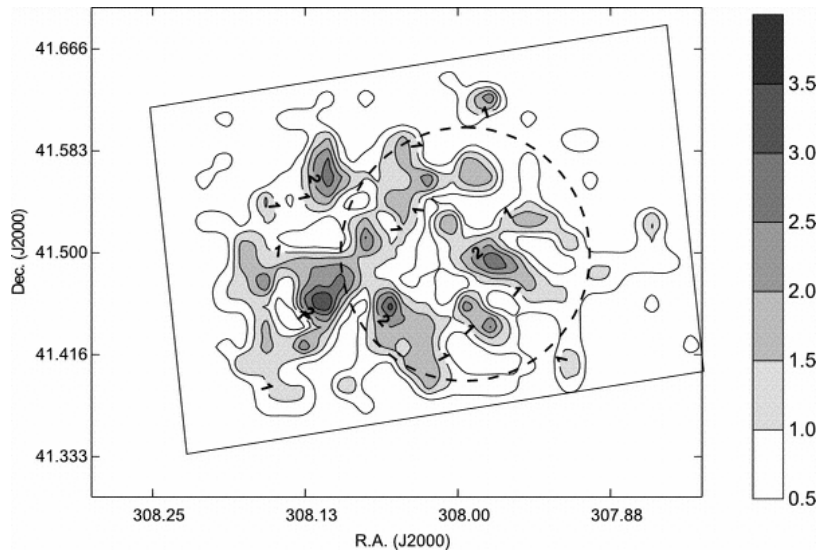

**Fig 1 – From Butt et al., 2006**. Surface density plot of the 240 pointlike X-ray sources detected in our Chandra observation of TeV J2032+4130. The cell size used for the smoothing was 1′ × 1′. The gray-scale bar at right indicates the number of pointlike X-ray sources $arcmin^{-2}$. The dotted circle shows the 1 σ extent of TeV J2032+4130, and the slanted inner rectangular region outlines the Chandra ACIS-I field of view. The X-ray point-source distribution is far from uniform: distinct concentrations of sources within the central ~7 arcmin of the field of view are revealed. Taken together, these local maxima are consistent with the size and location of TeV J2032+4130.

To make an association between these point-like sources and the TeV emission will require at least three additional *CHANDRA* observations of immediately adjacent fields to that shown in Fig 1. The mosaic of these four *CHANDRA* fields-of-view will then allow us to determine whether a true overdensity of point-like X-ray sources exists at the TeV source location.

The reported possible diffuse X-ray emission of intensity ~7 ×$10^{-13}$ ergs $cm^{-2}$ $sec^{-1}$ (2-10 keV) [Horns et al., 2007] from XMM, if real, would have been detected in our 50 ksec *CHANDRA* observation, but was not.